**Exploring the position of cities in global corporate research and development: a bibliometric analysis by two different geographical approaches**

**György Csomós**
Department of Civil Engineering, University of Debrecen
E-mail: csomos@eng.unideb.hu

**Géza Tóth**
Hungarian Central Statistical Office
E-mail: geza.toth@ksh.hu

**Abstract**

Global cities are defined, on the one hand, as the major command and control centres of the world economy and, on the other hand, as the most significant sites of the production of innovation. As command and control centres, they are home to the headquarters of the most powerful MNCs of the global economy, while as sites for the production of innovation they are supposed to be the most important sites of corporate research and development (R&D) activities. In this paper, we conduct a bibliometric analysis of the data located in the Scopus and Forbes 2000 databases to reveal the correlation between the characteristics of the above global city definitions. We explore which cities are the major control points of the global corporate R&D (home city approach), and which cities are the most important sites of corporate R&D activities (host city approach). According to the home city approach we assign articles produced by companies to cities where the decision-making headquarters are located (i.e. to cities that control the companies' R&D activities), while according to the host city approach we assign articles to cities where the R&D activities are actually conducted. Given Sassen's global city concept, we expect global cities to be both the leading home cities and host cities.

The results show that, in accordance with the global city concept, Tokyo, New York, London and Paris surpass other cities as command points of global corporate R&D (having 42 percent of companies' scientific articles). However, as sites of corporate R&D activities to be conducted, New York and Tokyo form a unique category (having 28 percent of the articles). The gap between San Jose and Boston, and the global cities has consistently narrowed because the formers are the leading centres of the fastest growing innovative industries (e.g. information technology and biotechnology) in the world economy, and important sites of international R&D activities within these industries. The emerging economies are singularly represented by Beijing; however, the position of Chinese capital (i.e. the number of its companies' scientific articles), has been strengthening rapidly.

**Keywords**: global city, multinational company, corporate research and development, scientific article

**1 Introduction**

Globalization and the spatial restructuring of the world economy have increased since the 1970s and can primarily be characterized by the expansion of trade, the growing volume of foreign direct investments (FDI), and the emergence of the new international division of labour (Fröbel et al. 1980, Cohen 1981). These developments have dramatically enhanced the developing countries' participation in the world economy. In the process of economic globalization, multinational companies (MNCs) have become the central orchestrators of a global reallocation of manufacturing away from core industrial countries towards the developing countries (Schoenberger 1988, Dicken 2007). MNCs interconnect nation-states, regions, and cities, and they exercise significant control over nation-states (Bonanno and Constance 2008). In this new world-system, cities have gradually become more important while the significance of nation-states has lessened (see, for example, Knox 1995; Scott et al. 2001; Sassen 2001; Sassen 2006). Alderson and Beckfield (2004: 812) argue that '*developments of the past few decades are seen as producing a new global hierarchy of cities, at the apex of which are located what have variously been referred to as "world cities" or "global cities." Such cities constitute the key nodes or command points that exercise power over other cities in a system of cities and, thus, the world economy*'. In her seminal work entitled, *The Global City,* Saskia Sassen (1991) specified New York, London, Tokyo, Frankfurt, and Paris as the leading examples of global cities. Furthermore, she defined the most important characteristics of global cities (Sassen 2001: 4):

*Beyond their long history as centers for international trade and banking, these cities now function in four new ways: first, as highly concentrated command points in the organization of the world economy; second, as key locations for finance and for specialized service firms, which have replaced manufacturing as the leading economic sector; third, as sites of production, including the production of innovation, in these leading industries; and fourth, as markets for the products and innovations produced.*

We highlight two important points concerning this definition: On the one hand, global cities are the outstanding command and control centres of the world economy, and on the other, they are the most significant sites for the production of innovation (Sassen 2001). The correlation between these two characteristics is the starting-point of this paper, and our main aim is to examine whether our theory is correct or not. Based on the characteristics of the global cities, we proposed a hypothesis, which needs to be confirmed by conducting a bibliometric analysis.

*Hypothesis*: Global cities are the major command and control centres of the world economy, and they are the most significant sites of the production of innovation[1]. As command and control centres, they are home to the headquarters of the most powerful MNCs of the global economy (Godfrey and Zhou 1999; Alderson and Beckfield 2004; Taylor et al. 2009; Taylor and Csomós 2012; Csomós 2013; Csomós and Tóth 2015). MNCs are often considered to be the most visible symbols of globalisation (Gavin 2001), because, for example, they have worldwide networks of subsidiaries, branch offices, customer service offices, and corporate research centres. To ensure the global competitiveness of firms, MNCs need to be highly involved in research and development (R&D) (Kogut and Zander 1993; Malecki 1997; Roth et al. 2009; Crespo et al. 2014). MNCs, wherever they are headquartered, tend to locate their R&D-oriented subsidiaries and corporate research centres into the most innovative environments in the world. Thus, if global cities are the major command and control centres and the most significant sites of the production of innovation in the world, they are home to not only the headquarters of the leading MNCs, but also host their R&D-oriented subsidiaries and corporate research centres. This means that, on the one hand, global cities are the major control points of corporate R&D (home city approach) and, on the other hand, the sites of international R&D activities (host city approach).

In this paper we put the above hypothesis to the test by conducting a bibliometric analysis. The intensity of corporate R&D activities can be measured through the number of patents and/or the number of patent citations (Santangelo 2002; Liu et al. 2006; Ribeiro et al. 2010; Ács 2011; Wang et al. 2011; Chang et al. 2012; Ribeiro et al. 2014; Wong and Wang 2015); the amount of R&D expenditures (Granstrand 1999; Kumar 2001; Yoo and Moon 2006; Piergiovanni and Santarelli 2013); the quantity and quality of research cooperation between companies and universities (Ramos-Vielba et al. 2010; Gao et al. 2011; Kneller et al. 2014; Leydesdorff et al. 2014; Feng et al. 2015), and it can be quantified by the number of scientific articles authored or co-authored by researchers from the companies (Hicks et al. 1994; Hicks 1995; Hullmann and Meyer 2003; Tijssen 2004; Furukawa and Goto 2006; Chang 2014). Several MNCs, especially those that operate in high-technology industries, are exceedingly involved in R&D activities; likewise, their researchers produce many scientific articles (Godin 1996; Chang 2014). Depending on the complexity of the MNC's organization and the geographical location of its R&D-oriented subsidiaries and corporate research centres, scientific articles can come from a number of domestic and foreign cities (Archibugi and Michie 1995; Cantwell 1995; Archibugi and Iammarino 1999; Gerybadze and Reger 1999; Santangelo 2000; Nam and Barnett 2011). Scientific articles published by researchers of MNCs can be examined by two different geographical approaches:

1. Home city approach: articles produced by companies can be assigned to cities where the decision-making headquarters are located (i.e. to cities that command the companies' R&D activities).
2. Host city approach: articles can be assigned to cities where the R&D-oriented subsidiaries and corporate research centres are located (i.e. to cities where the R&D activities are conducted).

---

[1] Saskia Sassen (1991) argues that global cities have become the most significant sites of the production of innovation. Of course, it is possible to achieve innovation without conducting R&D activities. This means that companies can be innovative without conducting R&D activities but by purchasing technology in the market through R&D contracting, licensing of technology and know-how, contracting technical and engineering services, and acquisition of machinery and equipment related to innovation (Veugelers 1997; Veugelers and Cassiman 1999). However, the phenomenon of "production of innovation", as to be mentioned by Sassen, is not equivalent to the phenomenon of "purchasing of innovation", because the former requires conducting advanced R&D activities, while the latter primarily requires money to buy innovation. Therefore, there is a close connection between R&D activities conducted by companies and the innovation produced by them.

The number of scientific articles assigned to the headquarters of the MNCs represents the power of the headquarters' cities as command centres of their global corporate R&D activities. The number of the scientific articles assigned to the locations of R&D-oriented subsidiaries, corporate research centres, and of course, headquarters of parent companies involved in R&D activities represents the power of these cities as sites of global corporate R&D activities. Of course, most cities are both headquarters of MNCs and locations of subsidiaries and corporate research centres, which means that they are not only command points for the companies' R&D activities, but locations where research activities are conducted. However, according to our hypothesis, global cities, i.e. New York, Tokyo, London, and Paris, are at the top of the hierarchy regarding both approaches; that is, they are supposed to have the largest number of scientific articles as headquarters' cities (home city approach) and as sites of R&D activities (host city approach) as well.

The structure of the paper is as follows: first, we demonstrate the theoritical background of the research, then in the empirical analysis of the paper, after introducing the data and methodology, we present the general and specific results of the research, ranking the cities on the basis of both approaches. Finally, we draw conclusions and point out future directions for study.

**2 Theoretical background of the research**

**2.1 Position of R&D-oriented subsidiaries in the organization of companies**

Multinational corporations are the main drivers for the internationalization of innovation activities (see, for example, Cantwell 1995; Gerybadze and Reger 1999; Narula and Zanfei 2004). Based on their R&D capacity, leading MNCs belong to the most significant research organizations in the world, which is represented by their huge amount of patents and scientific articles, as well as their extremely large R&D expenditure[2]. Recently, MNCs have become very complex organizations regarding their R&D activities. According to Santangelo (2000: 275), "*the shift towards a knowledge-based economy involves a shift in organisation away from top-down hierarchical infrastructures to flatter structures based on intra-firm networks of semi-autonomous corporate subsidiaries.*" In the 21th century's post-industrial knowledge-based economy, those companies have become the most competitive business organizations that consider knowledge to be their most strategically important resource (Kogut and Zander 1993; Roth et al. 2009; Crespo et al. 2014). Porter & van der Linde (1995) indicate that international competitiveness is based on innovation; therefore, MNCs tend to locate their R&D facilities into the most innovative environment in order to improve their organization's ability to leverage knowledge (Gerybadze and Reger 1999; Pearce 1999; Zander 1999; D'Agostino and Santangelo 2012). Adenfelt & Lagerström (2006: 382) claim that "*the differentiated MNC is more favourably positioned for leveraging of knowledge than the non-differentiated MNC, simply because of its access to more knowledge networks; both internal and external.*" According to Crespo et al. (2014), subsidiaries serve as key knowledge nodes capable of acquiring, converting, and transferring knowledge throughout the MNC. While Michailova and Mustaffa (2012) assert that subsidiaries are increasingly acknowledged as sources of knowledge both for the headquarters and for the peer subsidiaries. Nevertheless, even functionally different subsidiaries compete within the organization of the MNC to gain resources (Birkinshaw, 1996;

---

[2] In 2011, in terms of R&D expenditure, Johns Hopkins University was ranked first among the higher education institutions in the United States spending $ 2.1 billion on R&D activities, however, Johns Hopkins only ranked 53rd in the corporate R&D ranking of 2011. Moreover, the R&D expenditures of the leading ten multinational companies exceeded that of the United States' whole higher education sector (National Science Foundation 2012; EU Economics of Industrial Research & Innovation 2012).

Birkinshaw & Hood, 1998; Mudambi & Navarra, 2004). Mudambi et al. (2014: 109) indicate that "*a subsidiary that controls technology-focused resources and the technology-related function of the MNC's value chain gain strategic power in the firm, while subsidiaries that control business-related resources exercise decision-making power over the business-related functions of the value chain but they do not influence the strategic direction of the firm.*"

Concluding, we can state that globally competitive MNCs consider innovation and knowledge as vitally important to the success of the company and look to establish a network of competitive R&D-oriented subsidiaries within their organizations. In the era of the post-industrial knowledge-based economy, those cities have become important economic nodes that successfully attract innovative companies and their R&D-oriented subsidiaries and corporate research centres. According to Sassen's (2001) global city concept, these cities are the global cities as outstanding sites of the production of innovation. To prove this assumption, we conducted a bibliometric analysis of scientific articles published by researchers from leading multinational companies.

**2.2 Scientific articles reflecting on the R&D capacity of companies**

Measuring the scientific performance, innovation and R&D capacity, and knowledge transfer intensity of companies through the number and/or citation data of scientific articles is only matter of methodology, a form of bibliometric analysis (see, for example, Halperin and Chakrabarti 1987; Hicks et al. 1994; Godin 1996; Tijssen 2004; Tijssen and van Leeuwen 2006; Calero et al. 2007; Han 2007; Wong and Goh 2010; OECD 2011; Chang 2014; Abramo and D'Angelo 2015, among others). Furthermore, the motivations of corporate researchers when publishing scientific articles in academic journals is frequently studied; whether the individual's interest is to create articles to meet the business interest of companies; how the purpose of patent application influences the scientific content of articles, and whether companies consider their R&D activities to be a marketing message to attract outside researchers (Narin et al. 1987; Rosenberg 1990; Hicks 1995; Godin 1996; Kinney et al. 2004; Archambault and Larivière 2011; Li et al. 2015). However, regarding the aims of this analysis, it is more important to understand when the process of creating scientific articles appears among the phases of the innovation chain. The international competitiveness of world leading companies is fundamentally influenced by their innovation performance (Kafouros 2008). Applied research is essential for companies to invent new patentable technologies that contribute to the profitability of the company. Patented technologies are the catalyst of such new products and efficient services that increase the revenues, profits, and market values of companies (Jaffe 1986; Narin et al. 1987). Moreover, they help monopolize the position of companies in the global markets, which is subsequently translated into economic benefits (Archambault and Larivière 2011). While the importance and necessity of applied research is favourably recognized by the companies, basic research has an uncertain position in the corporate R&D portfolio. According to Rosenberg (1990: 165), companies "*will spend their own money on basic research only when they are reasonably confident that it will yield a rate of return on this investment in the generation of knowledge that is at least comparable to the rate of return that they would expect in some other form of investment in more tangible capital.*" For profit-oriented business organizations, the long-term return on basic research investments is a crucial factor because it may risk the profitability of the company. Thus, such companies can afford to carry out a wide range of basic research that has strong positions in the market. Narin et al. (1987) indicate that the main goal of R&D activities in the product life cycle is to produce scientific innovations (which result in scientific publications), which then lead to technological innovations (which result in patents).

Furthermore, there is a very important reason why corporate researchers are encouraged to publish their works in scientific journals. It is necessary to mention that most companies, especially those that are involved in high-tech industries, have a comprehensive intellectual property (IP) strategy as part of their corporate strategy. One important component of IP strategy is defensive publishing, which is an increasingly common tactic for protecting intellectual property. Defensive publishing is generally used by companies to disclose an invention to the public preventing other companies from claiming patents in the same area (Paasi et al., 2012.). According to Barrett (2002) successful defensive publication renders the competitor's invention obvious or lacking in novelty. Regarding the fact that in some industries competition between companies has become even more aggressive the phenomenon of defensive publishing significantly contributes to the increase of the number of articles published by corporate researchers.

In conclusion, it can be asserted that the opportunity of creating scientific articles by corporate researchers primarily belongs to those innovation-oriented powerful multinational companies that have strong basic research components in their R&D portfolio, and to companies for whom defensive publishing is a key component of their IP strategy. A number of studies demonstrate that these multinational companies operate in high-technology industries (e.g. aerospace and defence, biotechnology, communication technology, electronics, information technology, pharmaceuticals), and in some traditional industries that are highly involved in basic research (e.g. conglomerates, and different types of chemical industries) (Halperin and Chakrabarti 1987; Rosenberg 1990; Hicks et al. 1994; The New York Times 2002; Chang 2014).

## 3 Data and methodology

### 3.1 Data collection

#### 3.1.1 Multinational companies and industry classification

The world's largest multinational companies are ranked annually by Fortune 500 and Forbes 2000. In this analysis, we focus on the Forbes 2000 database for 2014, which ranks the world largest 2000 market-listed companies on the basis of a complex index that combines revenues, market values, assets and profits. Forbes classifies these companies into 81 industries; however, the global economic significance of these industries can be regarded as very different. For example, 223 companies are classified into the Regional Banks industry while the Leisure Products industry contains only one. Forbes' industry taxonomy roughly corresponds to that of the Global Industry Classification Standard (GICS) developed by MSCI Inc. and Standard & Poor's. The GICS classifies industries into ten industry sectors: Consumer Discretionary, Consumer Staples, Energy, Financials, Health Care, Industrials, Information Technology, Materials, Telecommunication Services, and Utilities. In this analysis, we classify the 81 Forbes' industries into the ten GICS industry sectors to reduce the differences between industries.

#### 3.1.2 Data source of the scientific articles

In this paper, we conducted a bibliometric analysis of data in the Scopus database. Scopus is the largest abstract and citation database, including almost 22,000 titles (20,800 peer-reviewed journals, 367 trade publications, and more than 400 book series) from 5000 publishers, and 6.4 million conference papers (Elsevier, 2014). Scopus offers the broadest, most integrated coverage available of scientific, technical, medical and social sciences including arts & humanities literature. A disadvantage of Scopus is that it does not have complete citation information for articles published before 1996 (Hengl et al. 2009; Vieira and Gomes 2009). However, considering the purpose of this analysis, Scopus is the most appropriate database; this is because it not only contains information on bibliometric data of parent companies, but also takes into account every single component of the company's organization (headquarters of parent companies, subsidiaries, corporate research centres, and branch offices). As of March 31, 2015, researchers employed by Forbes 2000 companies had 1,434,444 scientific articles (i.e. journal articles, conference papers, and book chapters) in Scopus. Regarding their bibliometric characteristics, the Forbes 2000 companies can be classified into four groups:
1) 1215 companies, especially companies involved in Financials, do not have any scientific articles in Scopus.
2) Several high-tech manufacturing companies and large industrial conglomerates have a globally significant network of subsidiaries and corporate research centres; therefore, the geographical distribution of their articles is very wide. For example, only 30 percent of the Swiss-based ABB industrial conglomerate's scientific articles belong to its Zurich headquarters, while 70 percent of the articles come from 29 subsidiaries and research centres in 25 cities on five continents. The Forbes 2000 companies control 1186 subsidiaries, branches, and corporate research centres of which researchers have scientific articles in Scopus.

3) Some companies, especially those that are involved in Financials, Utilities, and Consumer Staples, R&D activities are carried out in their headquarters only. For example, for companies involved in Financials, the ratio of headquarters/subsidiaries with articles in Scopus is 1:0.05 (Utilities: 1:0.23; Consumer Staples: 1:0.25).Which means that every 20$^{th}$ company has a subsidiary that has scientific articles, while the ratio for companies involved in Information Technology is 1:2.24 (Health Care: 1: 2.07; Materials: 1: 0.97).

4) The fewest number of companies R&D activities are carried out in only one city which is not home to its headquarters. This phenomenon is especially characteristic of companies that relocate their headquarters abroad for several reasons (usually due to tax optimization); however, with the exception of the companies' management, all other activities, even manufacturing and R&D, remain in the former headquarters city. For example, the Cleveland, Ohio-based industrial conglomerate, Eaton Corporation relocated its headquarters to Dublin, Ireland in 2012, while the company's R&D activities continue to be carried out in Cleveland (i.e. its scientific articles in Scopus originate from Cleveland).

**3.1.3 Territorial demarcation**

In light of the world city approach (see, Hall 1966; Friedmann 1986), we organized individual cities and towns into larger metropolitan areas. Florida and Jonas (1991), Lyons and Salmon (1995) and Sassen (2006) all stress that in the United States (and later in Western Europe), the relocation of headquarters from large cities to suburban areas (i.e. small cities, towns, villages) has become quite common since the 1970s. This phenomenon is what Florida and Jonas (1991) refer to as the decentralization of corporate organizations. According to Garreau (1991), Brenner (2002), and Ross and Levine (2012), this development has resulted in a significant change in the suburban network of metropolitan areas: suburbs began to have an active role in the economy, even in R&D activities. This latter process has also been highlighted by Grossetti et al. (2014) who examine the spatial distribution of scientific activities in countries worldwide considering metropolitan areas to be the basic territorial units of their analysis.

The New York metropolitan area is the best example to illustrate this phenomenon: New York City is the most significant headquarters city in the New York metropolitan area (63 percent of the companies headquartered in New York City). However, American and foreign Forbes 2000 companies settle headquarters, subsidiaries, branch offices, and corporate research centres in 47 smaller cities in the metropolitan area.

Metropolitan areas in the world are usually demarcated by national statistical offices, Metropolitan Statistical Areas in the United Sates are defined by the Office of Management and Budget, and Functional Urban Areas are defined by the ESPON project in the European Union (www.espon.eu). We assigned the Forbes 2000 companies and their 1186 subsidiaries, branches, and corporate research centres to 520 metropolitan areas.

**3.2 Methodology**

We approached this analysis from two viewpoints, which resulted in two different geographical aspects of corporate R&D activities. First, we assigned all scientific articles created at the company's headquarters, subsidiaries, branch offices and corporate research centres to the headquarters city; that is, to the city from which the company's whole organization is controlled. Thus, for example, all scientific articles of the Dutch conglomerate, Royal Philips, were assigned to its headquarters city, Amsterdam. However, less than 6 percent of the articles came from its headquarters. Second, we assigned the company's scientific articles to cities where its R&D activities were conducted, that is, to cities in which the articles were created. Quoting the above example, most of Royal Philips' articles came from 24 subsidiaries located in 19 cities in ten countries; moreover, 60 percent of them were created in Philips Research Eindhoven. These two approaches may lead to extreme variations: on the one hand, in spite of being the home of many corporate research centres or R&D-oriented subsidiaries, certain headquarters cities can have a large number of scientific articles (as if they were important research nodes), and on the other, in spite of being the headquarters of many global MNCs, some cities can have many scientific articles because they are significant sites of corporate R&D activities. Of course, several world cities, especially the largest ones, such as New York, Tokyo, London, and Paris, act as both important command and control centres in the world economy (Csomós 2013), and as sites of subsidiaries and corporate research centres of domestic and foreign MNCs (Lyons and Salmon 1995; Alderson and Beckfield 2012). This means that the largest global cities are supposed to be at the top of the hierarchy by having the largest number of scientific articles by either approach.

## 4 Results

### 4.1 General results

#### 4.1.1 Correlation between industry sectors and scientific articles

The largest industry sector of Forbes in terms of the number of the companies is the Financials, comprising 30 percent of all companies (Table 1). The combined number of companies of the second and third largest sectors, Industrials and Consumer Discretionary, do not reach that of the Financials. However, considering the number of companies that have scientific articles in Scopus, the Industrials sector is ranked top thanks to the industrial conglomerates and their large network of subsidiaries. It can be seen in Table 1 that despite having less than one fifth and one third companies than the Financials and the Industrials, the Information Technology sector is in the leading position in terms of the number of the scientific articles and is ranked second regarding their per-company values. Because companies involved in the Health Care sector, especially in the Pharmaceuticals industry, have the second largest number of scientific articles in Scopus, even though their number is very few in Forbes (91), it is not surprising that they have exceptional per-company values (6784 articles per company). As has been mentioned previously in terms of the number of Forbes 2000 companies, the Financials sector ranks top; however, all companies in the sector have fewer articles in Scopus than that of the Eastman Kodak (47[th] in the corporate ranking).

Table 1. Scientific articles of Forbes 2000 companies by industry sectors

| GICS Sector | No. of MNCs | No. of subsidiaries, and R&D centres | Total | No. of articles | Articles / MNC | Articles / Total |
|---|---|---|---|---|---|---|
| Consumer Discretionary | 247 | 76 | 323 | 97,389 | 394 | 302 |
| Consumer Staples | 153 | 39 | 192 | 38,891 | 254 | 203 |
| Energy | 126 | 68 | 194 | 102,001 | 810 | 526 |
| Financials | 587 | 27 | 614 | 8,695 | 15 | 14 |
| Health Care | 91 | 188 | 279 | 316,265 | 3,475 | 1,134 |
| Industrials | 325 | 312 | 637 | 297,167 | 914 | 467 |
| Information Technology | 105 | 235 | 340 | 339,102 | 3,230 | 997 |
| Materials | 194 | 189 | 383 | 142,769 | 736 | 373 |
| Telecommunication Services | 62 | 27 | 89 | 62,985 | 1,016 | 708 |
| Utilities | 110 | 25 | 135 | 29,180 | 265 | 216 |
| TOTAL | 2,000 | 1,186 | 3,186 | 1,434,444 | | |

In conclusion, we can state that those cities have the largest number of scientific articles that are home to companies primarily involved in Health Care, Information Technology, and Industrials sectors.

#### 4.1.2 Ranking companies by the number of scientific articles

The largest number of the Forbes 2000 companies' scientific articles comes from IBM, which accounts for six percent of all articles, and almost 25 percent of the articles of the Information Technology sector (Table 2). Scientific articles of IBM originate from 14 subsidiaries and corporate research centres; however, 70 percent of them belong to the Thomas J. Watson Research Center (Yorktown Heights, New York) and their headquarters (Armonk, New York). The second largest number of articles comes from Alcatel-Lucent. The company was formed in 2006 by the merger of Alcatel (France) and Lucent Technologies (United States), and because the acquisition was initiated by Alcatel, the common headquarters of the new company was relocated to Paris. Lucent Technologies brought to the merger the prestigious Bell Laboratories (Murray Hill, New Jersey), the largest research centre in the telecommunication industry, possessing more than 30 thousand patents at that time. However, the scientific superiority of the American part of Alcatel-Lucent is clearly represented by the fact that 83 percent of the company's articles come from Bell Labs (that is, it could be the second largest individual company in terms of the number of the articles) in contrast with the 10 percent share of the Paris headquarters.[3] Alcatel-Lucent is a typical example of how profitable multinational companies broaden their scientific portfolio by acquiring companies that have significant R&D activities (Bena and Li 2014).

Table 2. Leading companies by the number of scientific articles

| Rank | MNC | Country | Industry | No. of articles | Percentage within the sector |
|---|---|---|---|---|---|
| 1 | IBM | United States | Information Technology | 83,669 | 24.67 |
| 2 | Alcatel-Lucent | France | Information Technology | 61,501 | 18.14 |
| 3 | Pfizer | United States | Health Care | 43,302 | 13.69 |
| 4 | GlaxoSmithKline | United Kingdom | Health Care | 36,743 | 11.62 |
| 5 | NTT | Japan | Telecommunication Services | 35,106 | 55.74 |
| 6 | General Electric | United States | Industrials | 29,900 | 10.06 |
| 7 | Novartis | Switzerland | Health Care | 29,024 | 9.18 |
| 8 | Hitachi | Japan | Industrials | 28,719 | 9.66 |
| 9 | Merck & Co | United States | Health Care | 27,928 | 8.83 |
| 10 | Royal Philips | Netherlands | Industrials | 27,553 | 9.27 |
| 11 | Roche Holding | Switzerland | Health Care | 27,421 | 8.67 |
| 12 | Siemens | Germany | Industrials | 25,768 | 8.67 |
| 13 | E. I. du Pont de Nemours | United States | Materials | 20,646 | 14.46 |
| 14 | Microsoft | United States | Information Technology | 20,610 | 6.08 |
| 15 | Bayer | Germany | Materials | 20,157 | 14.12 |
| 16 | NEC | Japan | Industrials | 19,347 | 6.51 |
| 17 | Intel | United States | Information Technology | 18,205 | 5.37 |
| 18 | Eli Lilly & Co | United States | Health Care | 17,759 | 5.62 |
| 19 | Sanofi | France | Health Care | 17,570 | 5.56 |
| 20 | Hewlett-Packard | United States | Information Technology | 17,256 | 5.09 |
| 21 | Toshiba | Japan | Industrials | 17,219 | 5.79 |
| 22 | Exxon Mobil | United States | Energy | 15,561 | 15.26 |
| 23 | Bristol-Myers Squibb | United States | Health Care | 15,357 | 4.86 |
| 24 | AstraZeneca | United Kingdom | Health Care | 15,201 | 4.81 |
| 25 | Boeing | United States | Industrials | 14,734 | 4.96 |

The Industrial sector's largest companies in terms of the number of scientific articles are conglomerates (General Electric, Royal Philips, and Siemens) and electronics companies (Hitachi, NEC, and Toshiba). This is due to the fact that companies belonging to both types of companies have similar production structures. General Electric (GE) is the largest industrial conglomerate in the world; it operates in some low-technology industries (e.g. mining) as well as innovation-oriented high-technology industries (e.g. aerospace, health care). Moreover, 25 percent of the company's revenue is generated by financial businesses (GE 2014). It is not surprising that the majority of the conglomerates and electronics companies' scientific articles (e.g. 90 percent of GE's articles) come from their high-technology segment. Because these companies have complex production structures, the scientific subject areas of their articles also tend to be very broad.

As can be seen in Table 2, one-third of the leading companies are involved in the Health Care sector. It is common for these companies to be part of the Pharmaceuticals industry, of which, companies contribute to less than 50 percent of the total number of the companies in the sector but have 88 percent of all articles. The largest non-Pharmaceuticals company is the American biotechnology firm, Amgen, standing in 11th position. In the ranking of leading companies, Pfizer (United States) and GlaxoSmithKline (United Kingdom) are in 3rd and 4th positions. These pharmaceutical companies' share, in terms of the number of scientific articles in the Health Care sector, is 25 percent. In contrast to companies involved in other industries, a specific characteristic of pharmaceutical companies is that the volume of their R&D activities is extremely high (Calero et al. 2007); that is, many researchers at numerous pharmaceutical companies create many scientific articles. Furthermore, the Health Care sector, especially the Pharmaceuticals industry, is characterized by numerous relocations, which is a natural consequence of mergers and acquisitions, and the effort of multinational companies to optimize their tax burdens.

---

[3] The Finnish-based Nokia announced that it would acquire Alcatel-Lucent for $16.6 billion in 2015 to become the leading network equipment provider in the world, surpassing its Swedish rival Ericsson (Forbes 2015). The headquarters of the company will most likely remain in Espoo (Helsinki metropolitan area); that is, Paris will lose its significant position as command and control centre of corporate R&D activities, while the position of Murray Hill (New York metropolitan area) will not change.

Nippon Telegraph & Telephone (NTT) is the most dominant company in its industry sector, having 56% of the total number of articles in the Telecommunication Services sector. The main reason for this is that the innovation potential of NTT is served by the largest network of research institutes in the industry.[4] While the Japanese NTT is 5th in the global ranking of leading companies, having more than 35 thousand articles, its rival in the sector, the French-based, Orange (formerly France Télécom), has 7100 articles and is only 54th.

Despite the fact that Financials is the largest industry sector in terms of the number of Forbes 2000 companies, it has the fewest number of scientific articles in Scopus. The American real estate investment trust, Weyerhaeuser, tops the Financials sector (210th in the global ranking) having 833 articles. However, the company is involved not only in financial businesses, but is also one of the world's largest private owners of timberlands (Sun et al. 2013). As such, 90 percent of its articles derive from the subject areas of Agricultural and Biological Sciences, Engineering, Environmental Sciences, and Material Sciences. Excluding Weyerhaeuser, the German insurance company, Allianz (223 articles, 378th position), and the Bank of Greece (162 articles, 418th position) are in key leading positions.

**4.1.3 Spatial classification of articles of MNCs by home countries and host countries**

The spatial classification of articles produced by MNCs can be investigated on the basis of two approaches: 1) assigning articles to home countries, i.e. to countries that are home to headquarters of MNCs; 2) assigning articles to host countries, i.e. to countries that host subsidiaries, R&D centres and headquarters where research is conducted.

Table 3. The number of articles assigned to both home countries and host countries

| Rank | Country | No. of MNCs | No. of articles assigned to home countries | Rank | Country | No. of subsidiaries, and R&D centres | No. of articles assigned to host countries |
|---|---|---|---|---|---|---|---|
| 1 | United States | 565 | 611,532 | 1 | United States | 573 | 681,800 |
| 2 | Japan | 226 | 248,684 | 2 | Japan | 124 | 236,776 |
| 3 | France | 66 | 120,272 | 3 | Germany | 120 | 90,424 |
| 4 | Switzerland | 48 | 87,326 | 4 | United Kingdom | 78 | 85,686 |
| 5 | Germany | 52 | 86,686 | 5 | France | 54 | 71,762 |
| 6 | United Kingdom | 94 | 80,641 | 6 | Switzerland | 45 | 52,242 |
| 7 | Netherlands | 27 | 66,194 | 7 | Netherlands | 35 | 49,739 |
| 8 | South Korea | 61 | 30,019 | 8 | China | 66 | 31,595 |
| 9 | China | 207 | 25,194 | 9 | South Korea | 31 | 29,882 |
| 10 | Finland | 12 | 11,330 | 10 | Sweden | 20 | 15,671 |
| 11 | Denmark | 14 | 8,041 | 11 | Finland | 25 | 10,330 |
| 12 | Sweden | 26 | 7,570 | 12 | India | 63 | 9,935 |
| 13 | India | 53 | 6,937 | 13 | Denmark | 17 | 7,924 |
| 14 | Taiwan | 47 | 6,084 | 14 | Canada | 72 | 7,351 |
| 15 | Norway | 9 | 5,272 | 15 | Italy | 17 | 6,772 |

Most Forbes 2000 companies are headquartered in the United States and Japan; therefore, it is not surprising that the majority of the scientific articles come from these two countries (Table 3). As home countries, France and Switzerland are in a better position than as host countries because in the latter relation both Germany and the United Kingdom surpass them. The reason for this is that several France-based and Switzerland-based large MNCs tend to conduct their R&D activities outside of their own country: approximately 40-40 percent of both countries' scientific articles are created in their foreign subsidiaries and research labs. The United States hosts the largest number of foreign-owned (primarily Japanese and Western European) research-oriented subsidiaries and R&D centres, while Canada is the main target area of MNCs from the United States.

---

[4] NTT Basic Research Laboratories (http://www.brl.ntt.co.jp/E/introduction/introduction.html)

Among both the major home countries and host countries there are only two emerging economies: 60 percent of the scientific articles of developing countries come from China, and 20 percent of them from India.

**4.2 Ranking cities on the basis of the home city and the host city approaches**

**4.2.1 Home city approach: cities that command the MNCs' R&D activities**

Hall (1966), Friedmann (1986), and Sassen (2001) all stress that New York, London, Tokyo, and Paris are all included in the major global cities/world cities. Furthermore, according to Godfrey and Zhou 1999, Alderson and Beckfield 2004, Taylor and Csomós 2012, and Csomós 2013, these cities are the leading command and control centres of the global economy by acting as the headquarters of powerful MNCs. Therefore, global cities are also the most important command points of worldwide corporate R&D activities, which is supported by the fact that they have the largest number of scientific articles. It can be seen in Table 4 and Fig. 1 that Tokyo, New York, Paris, and London (especially the former two) excel in terms of the number of articles; they have 42 percent of all scientific articles, however only 18 percent of the Forbes 2000 companies are headquartered in them. Perhaps the leading position of the four global cities is not surprising considering that they comprise not only the majority of the most powerful multinational companies in the world but also many start-up companies involved in the fastest growing industries, such as nanotechnology, biotechnology, and information technology. Furthermore, companies that have been operating for a long time, primarily in the chemical industry, pharmaceuticals and electronics, significantly affect the number of articles in their headquarters cities. For example, the Philadelphia-based E. I. du Pont de Nemours, founded in 1802, has more than 20 thousands articles in Scopus, in contrast to Google Inc. (located in the Mountain View San Jose metropolitan area), which has 3750 articles in Scopus but has only been operating since 1998. However, researchers at DuPont published 342 articles in 2014 while researchers at Google created 516 articles, that is, the gap between these companies has been closing. The DuPont-Google example clearly illustrates how the scientific performance of companies involved in traditional and modern industries has been shifting. In the recent past, technological change has occurred at a rapid pace, due to fast-growing industries like nanotechnology, biotechnology, and information technology (Hullmann and Meyer 2003; Nicolini and Nozza 2008). For this reason, cities that host leading companies involved in modern industries (for example, San Jose and Boston) will have an increasing number of scientific articles and will overtake the global cities.

Table 4. Ranking home cities by the number of the scientific articles of MNCs

| Rank | Metros/Cities | Country | No. of MNCs | No. of articles assigned to home cities | Percentage within the dataset |
|---|---|---|---|---|---|
| 1 | Tokyo | Japan | 147 | 205,718 | 14.34 |
| 2 | New York | United States | 88 | 191,369 | 13.34 |
| 3 | Paris | France | 62 | 118,749 | 8.28 |
| 4 | London | United Kingdom | 76 | 80,293 | 5.60 |
| 5 | Basel | Switzerland | 7 | 59,405 | 4.14 |
| 6 | Amsterdam/Randstad | Netherlands | 22 | 59,187 | 4.13 |
| 7 | San Jose | United States | 28 | 56,330 | 3.93 |
| 8 | Bridgeport | United States | 12 | 40,919 | 2.85 |
| 9 | Chicago | United States | 34 | 36,914 | 2.57 |
| 10 | Dallas | United States | 22 | 35,281 | 2.46 |
| 11 | Munich | Germany | 9 | 34,239 | 2.39 |
| 12 | Osaka/Keihanshin | Japan | 31 | 33,406 | 2.33 |
| 13 | Cologne/Rhine-Ruhr | Germany | 15 | 29,868 | 2.08 |
| 14 | Washington | United States | 17 | 27,273 | 1.90 |
| 15 | Seoul | South Korea | 53 | 26,884 | 1.87 |
| 16 | Detroit | United States | 10 | 26,680 | 1.86 |
| 17 | Seattle | United States | 10 | 21,686 | 1.51 |
| 18 | Philadelphia | United States | 14 | 21,578 | 1.50 |
| 19 | Beijing | China | 53 | 20,863 | 1.45 |
| 20 | Indianapolis | United States | 4 | 17,849 | 1.24 |
| 21 | Boston | United States | 14 | 17,258 | 1.20 |
| 22 | Houston | United States | 28 | 15,118 | 1.05 |
| 23 | San Francisco | United States | 19 | 13,163 | 0.92 |
| 24 | Zurich | Switzerland | 25 | 12,630 | 0.88 |
| 25 | Helsinki | Finland | 12 | 11,330 | 0.79 |
| 26 | Geneva | Switzerland | 5 | 10,891 | 0.76 |
| 27 | Midland, Michigan | United States | 1 | 10,332 | 0.72 |
| 28 | Oxnard | United States | 1 | 9,493 | 0.66 |
| 29 | Mannheim/Rhine-Neckar | Germany | 5 | 9,429 | 0.66 |
| 30 | Rochester | United States | 3 | 8,680 | 0.61 |
| 31 | Minneapolis | United States | 16 | 8,630 | 0.60 |
| 32 | Copenhagen | Denmark | 10 | 7,897 | 0.55 |
| 33 | St. Louis | United States | 8 | 7,628 | 0.53 |
| 34 | Cincinnati | United States | 7 | 7,417 | 0.52 |
| 35 | Nagoya/Chūkyō | Japan | 16 | 6,347 | 0.44 |
| 36 | Stockholm | Sweden | 23 | 4,857 | 0.34 |
| 37 | Stavanger | Norway | 2 | 4,793 | 0.33 |
| 38 | Dublin | Ireland | 18 | 4,710 | 0.33 |
| 39 | Rome | Italy | 6 | 4,448 | 0.31 |
| 40 | Los Angeles | United States | 18 | 4,365 | 0.30 |
| 41 | Hartford | United States | 6 | 4,349 | 0.30 |
| 42 | Frankfurt/Rhine-Main | Germany | 8 | 4,317 | 0.30 |
| 43 | Stuttgart | Germany | 4 | 4,265 | 0.30 |
| 44 | Lausanne | Switzerland | 3 | 4,222 | 0.29 |
| 45 | Corning | United States | 1 | 4,010 | 0.28 |
| 46 | Heerlen/South Limburg | Netherlands | 2 | 3,888 | 0.27 |
| 47 | Rio de Janeiro | Brazil | 4 | 3,855 | 0.27 |
| 48 | Brussels | Belgium | 12 | 3,722 | 0.26 |
| 49 | Hannover–Wolfsburg | Germany | 5 | 3,562 | 0.25 |
| 50 | Mumbai | India | 21 | 3,513 | 0.24 |
| | Top 50 Metros/Cities | | 1,017 | 1,363,610 | 95.06 |
| | Total of 381 Metros/Cities | | 2,000 | 1,434,444 | 100.00 |

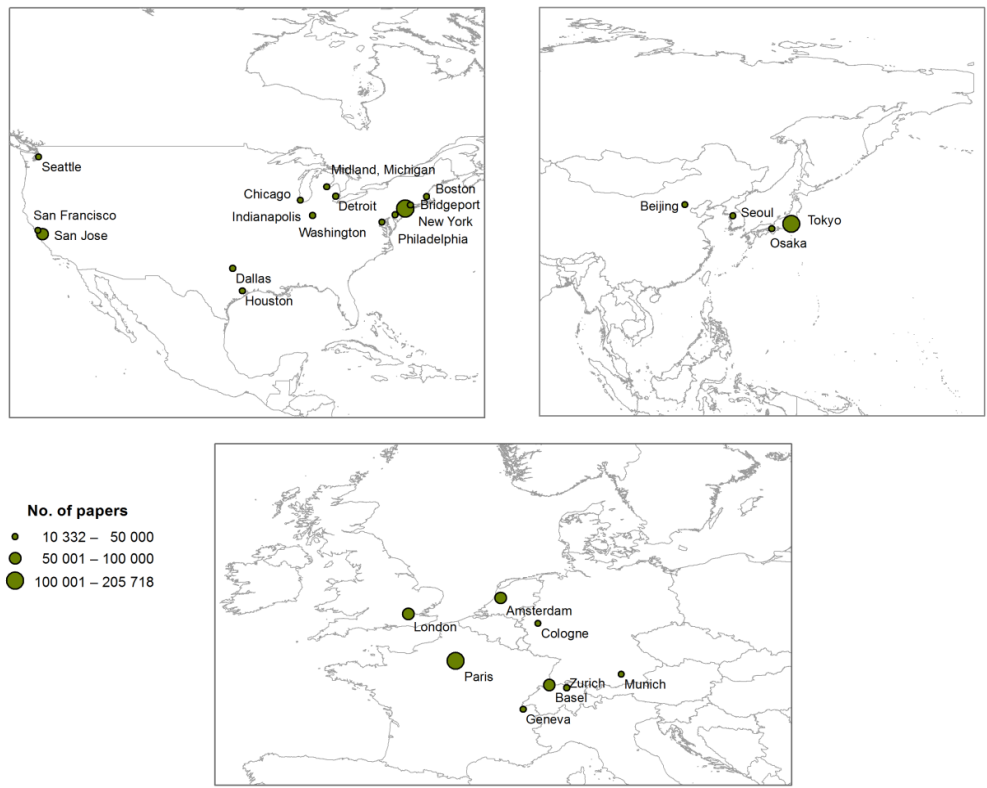

Fig. 1. Mapping the geographical location of home cities with more than 10 thousand articles

Headquarters cities with more than 10 thousand articles were located in the following well-defined geographical areas: 1) the East Coast of the United States, the surroundings of the Great Lakes, the Texas Triangle, and the San Francisco Bay Area; 2) the London-Paris-Amsterdam triangle and the Cologne-Munich-Basel-Zurich corridor in Europe; 3) East Asia where megacities, such as Tokyo, Osaka, Seoul, and Beijing create isolated islands.

## 4.2.2 Host city approach: cities where the R&D activities are conducted

It can be determined that there is clear overlap between the geographical location of the major command and control centres in the world economy and cities from which corporate R&D activities are controlled. However, the geographical location of cities where corporate R&D activities are conducted and where scientific articles are created (i.e. host cities of R&D activities) shows a far different pattern. The main reason for this is that MNCs organize their R&D activities in subsidiaries and corporate research centres while the headquarters generally remain responsible for only management activities. Companies tend to locate their R&D-oriented subsidiaries and research centres in the most innovative environment where they can gain knowledge and skills (Gerybadze and Reger 1999; Pearce 1999; Zander 1999; Fromhold-Eisebith 2002; D'Agostino and Santangelo 2012). Therefore, if a city ranks in the top as a global control centre of corporate R&D activities, it is not guaranteed that significant R&D activities are conducted there. The question is whether there are overlaps between the rankings created by the two approaches and what are the reasons for the differences if any.

From Table 4 and Table 5, it can be discerned that Tokyo, London, and Paris host significantly more headquarters than R&D-oriented subsidiaries and research centres while New York shows a balanced pattern. The main reason for this is that the majority of the largest financial companies (e.g. banks, insurance companies, real estate investment trusts) tend to be headquartered in the global cities. However, these companies' R&D activities are very poor, producing only a few scientific articles (see Section 3.1.1). Contrast this with cities like San Jose and Boston that host several R&D-oriented subsidiaries and research centres of companies headquartered in other cities, even in global cities; that is, these cities may have a special role in the international corporate R&D activities.

Table 5 shows that New York has the largest number of scientific articles and is the most significant site of corporate R&D activities in the world. Some 70 percent of New York's articles come from three companies' four research centres: Pfizer, two facilities of IBM, and Alcatel-Lucent's Bell Labs. New York is closely followed by Tokyo. These two cities excel in terms of the number of articles; in essence, making the world bipolar with regard to corporate R&D activities. Nevertheless, there is a significant difference between New York and Tokyo in that most articles created in New York come from subsidiaries and research centres owned by foreign companies (e.g. 25 percent from Bell Labs), while less than one percent of Tokyo's articles belong to foreign companies' subsidiaries. San Jose appears between the global cities, London and Paris. San Jose is the most important American and international centre of the information technology industry; what is more, several European and Japanese companies involved in information technology, pharmaceuticals, and electronics (for example, Hitachi, Roche Holding, Royal Philips) operate large research centres in the San Jose/Silicon Valley area. San Jose is one of the fastest growing sites of corporate R&D activities in the world, which is clearly represented by the fact that the largest number of its scientific articles was created in 2014.

London is in third position in the ranking, and it shows a very balanced pattern regarding both approaches: the difference between the numbers of its articles as a headquarters city and as a site of corporate R&D activities is less than one percent. Paris, on the contrary, as a site of corporate R&D activities dropped a position due to the fact that New York hosts the largest research centre owned by a French company (i.e. Alcatel-Lucent's Bell Labs). The difference between the numbers of Paris's articles with respect to the two approaches is more than 43 percent.

As can be seen in Table 5, Boston is in 6$^{th}$ position, while as a headquarters' city, it is only in 21$^{st}$ position (see, Table 4), and the difference between the numbers of its articles is 130 percent. According to Audretsch (1998: 18), much of the innovative activity is less associated with footloose multinational companies and is more closely associated with high-tech innovative regional clusters. Owen-Smith & Powell (2004: 9) assert that Boston is a strong candidate for a geographic region where information could diffuse widely and informally through a thriving technological community and labour market. Therefore, it is not surprising that such an innovation-oriented scientific and economic environment has emerged around research universities (for example, Harvard University and MIT) and high-tech start-up companies (Tödtling 1994) that make Boston attractive to American and foreign multinational companies.

Table 5. Ranking host cities by the number of scientific articles of companies, subsidiaries, R&D centres

| Rank | Metros/Cities | Country | No. of companies, subsidiaries, R&D centres | No. of articles assigned to cities where research was conducted | Percentage within the dataset |
|---|---|---|---|---|---|
| 1 | New York | United States | 88 | 209,377 | 14.60 |
| 2 | Tokyo | Japan | 90 | 194,773 | 13.58 |
| 3 | London | United Kingdom | 50 | 79,480 | 5.54 |
| 4 | San Jose | United States | 39 | 74,691 | 5.21 |
| 5 | Paris | France | 47 | 68,160 | 4.75 |
| 6 | Boston | United States | 25 | 39,556 | 2.76 |
| 7 | Chicago | United States | 25 | 35,361 | 2.47 |
| 8 | Osaka/Keihanshin | Japan | 18 | 31,351 | 2.19 |
| 9 | Basel | Switzerland | 13 | 30,434 | 2.12 |
| 10 | Munich | Germany | 21 | 28,429 | 1.98 |
| 11 | Detroit | United States | 12 | 27,601 | 1.92 |
| 12 | Seoul | South Korea | 28 | 26,747 | 1.86 |
| 13 | Dallas | United States | 13 | 26,157 | 1.82 |
| 14 | Philadelphia | United States | 20 | 25,243 | 1.76 |
| 15 | Cologne/Rhine-Ruhr | Germany | 29 | 24,360 | 1.70 |
| 16 | Beijing | China | 28 | 23,115 | 1.61 |
| 17 | Eindhoven | Netherlands | 2 | 19,492 | 1.36 |
| 18 | Seattle | United States | 12 | 18,604 | 1.30 |
| 19 | Indianapolis | United States | 5 | 17,976 | 1.25 |
| 20 | Amsterdam/Randstad | Netherlands | 19 | 17,095 | 1.19 |
| 21 | Bridgeport | United States | 6 | 16,625 | 1.16 |
| 22 | Washington | United States | 13 | 16,257 | 1.13 |
| 23 | San Francisco | United States | 18 | 13,858 | 0.97 |
| 24 | Houston | United States | 26 | 13,599 | 0.95 |
| 25 | Los Angeles | United States | 15 | 12,540 | 0.87 |
| 26 | Albany-Schenectady | United States | 2 | 10,805 | 0.75 |
| 27 | Zurich | Switzerland | 18 | 10,616 | 0.74 |
| 28 | Midland, Michigan | United States | 2 | 9,863 | 0.69 |
| 29 | Mannheim/Rhine-Neckar | Germany | 9 | 9,857 | 0.69 |
| 30 | Rochester | United States | 4 | 9,656 | 0.67 |
| 31 | Arnhem-Nijmegen | Netherlands | 2 | 9,625 | 0.67 |
| 32 | Trenton | United States | 7 | 9,450 | 0.66 |
| 33 | Helsinki | Finland | 19 | 8,975 | 0.63 |
| 34 | Cincinnati | United States | 7 | 8,557 | 0.60 |
| 35 | Oxnard | United States | 1 | 8,454 | 0.59 |
| 36 | Copenhagen | Denmark | 14 | 7,745 | 0.54 |
| 37 | Minneapolis | United States | 12 | 7,593 | 0.53 |
| 38 | St. Louis | United States | 9 | 7,552 | 0.53 |
| 39 | Nagoya/Chūkyō | Japan | 6 | 6,587 | 0.46 |
| 40 | Brussels | Belgium | 35 | 6,511 | 0.45 |
| 41 | Geneva | Switzerland | 6 | 6,500 | 0.45 |
| 42 | Stuttgart | Germany | 10 | 6,346 | 0.44 |
| 43 | Frankfurt/Rhine-Main | Germany | 19 | 6,331 | 0.44 |
| 44 | Pittsburgh | United States | 10 | 5,936 | 0.41 |
| 45 | Umea | Sweden | 1 | 5,360 | 0.37 |
| 46 | Stavanger | Norway | 1 | 4,790 | 0.33 |
| 47 | Rome | Italy | 7 | 4,580 | 0.32 |
| 48 | Berlin | Germany | 3 | 4,501 | 0.31 |
| 49 | Hamburg | Germany | 7 | 4,438 | 0.31 |
| 50 | Lausanne | Switzerland | 3 | 4,410 | 0.31 |
| | Top 50 Metros/Cities | | 876 | 1,275,919 | 88.95 |
| | Total of 360 Metros/Cities | | 2,000 | 1,434,444 | 100.00 |

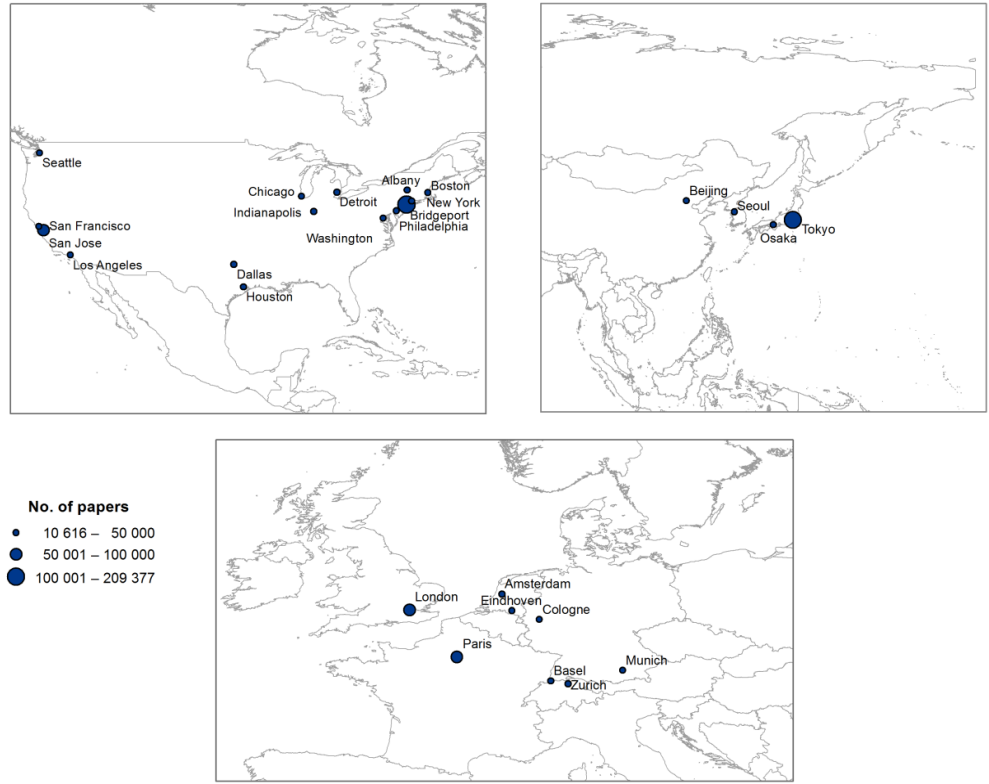

Fig. 2. Mapping the geographical location of host cities with more than 10 thousand articles

When comparing Table 4 and Table 5, it can be stated that there is a clear overlap between the geographical distribution of the locations of the major headquarters' cities and host cities of R&D activities. However, the most important difference is that European cities (especially Paris and Amsterdam/Randstad) have a much weaker position as sites of R&D compared to that of cities in the United States. Emerging economies are only represented by Beijing, in 16th position among the leading cities, followed by Rio de Janeiro in 56th position. By 2014, Beijing had become one of the most significant command and control centres in the world economy thanks to the massive growth of its financial sector. However, the position of Chinese capital as a site of corporate R&D activities is still poor, which is represented by the fact that, in spite of being a headquarters for 53 *Forbes* 2000 companies, 69 percent of its scientific articles come from three oil companies (PetroChina, Sinopec, and CNOOC). Nevertheless, by building a headquarters' economy (Pan et al. 2015), besides Chinese firms, Beijing attracts many foreign multinational companies (for example, IBM, Intel, NEC, and Toshiba), which locate not only their Chinese main offices to the capital city (see, for example, Wang, 2011), but also their R&D-oriented subsidiaries. For example, Beijing hosts the largest subsidiary of Microsoft in terms of the number of the articles.

### 4.2.3 Comparison of the home city approach and the host city approach

Considering the results of the home city and the host city approaches, we have classified cities into four categories. The classification is based on whether a given city has more or fewer articles as home cities as compared to their being host cities. Fig. 3 shows the four categories: 1) cities that have more than 10001 articles as host cities: 5 cities (indicated by dark red); 2) cities that have 1001-10000 articles as host cities: 31 cities (indicated by light red); 3) cities that have 1001-10000 articles as home cities: 17 cities (indicated by light blue); 4) cities that have more than 10001 articles as home cities: 6 cities (indicated by dark blue).

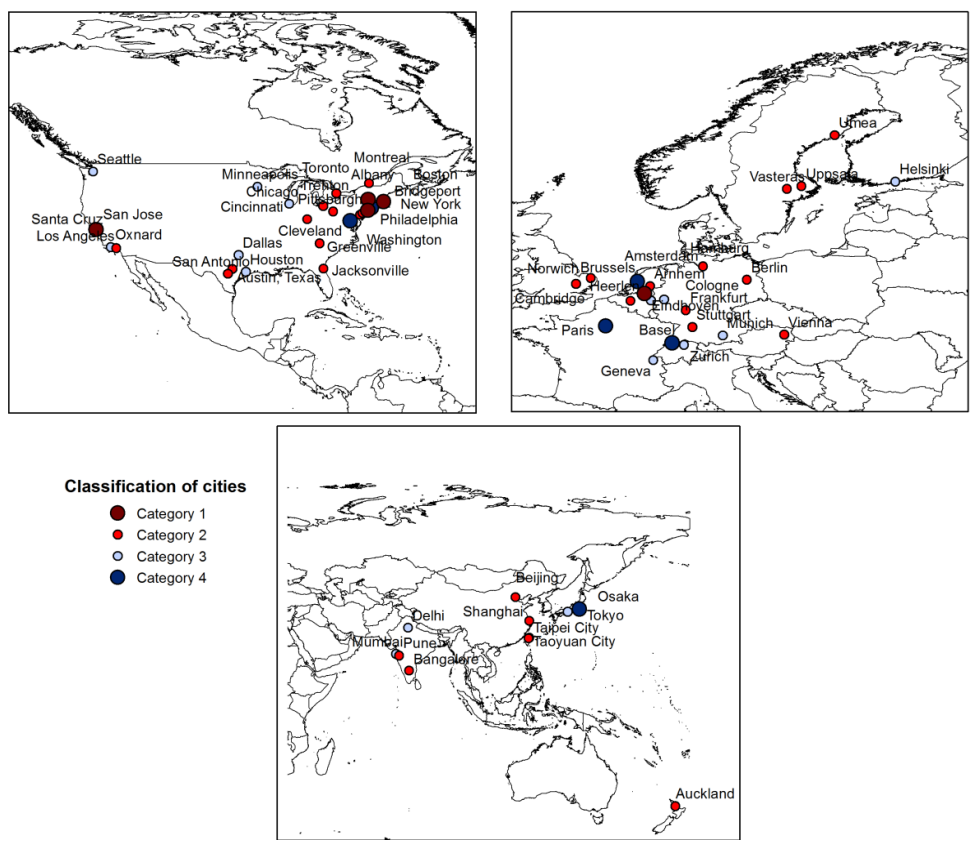

Fig. 3. Mapping cities on the basis of whether they have more or fewer articles as home cities as compared to their being host cities

As can be seen in Fig. 3, both Tokyo and Paris are included in the fourth category, that is, they have fewer articles as cities hosting corporate R&D than as home cities; however, in the case of Tokyo, the difference does not seem to be significant. New York is undoubtedly the most important city in the world for hosting the R&D activities of MNCs. A globally influential R&D area has emerged on the northeast coast of the United States with Boston and New York in the centre. Bridgeport is a significant home city (see Table 4) because it is home to General Electric (GE), the world's largest industrial conglomerate, however, its position as a city hosting corporate R&D is much weaker. The reason for this that numerous R&D-oriented GE subsidiaries and research centres operate worldwide, but the largest one in terms of the number of scientific articles is located in the New York metropolitan area strengthening the position of New York.

Table 6. Ranking cities on the basis of whether they have more or fewer articles as home cities as compared to their being host cities

| Rank | Host cities | Country | No. of articles | Home cities | Country | No. of articles |
|---|---|---|---|---|---|---|
| 1 | Boston | United States | 22,298 | Paris | France | 50,589 |
| 2 | San Jose | United States | 18,361 | Amsterdam | Netherlands | 42,092 |
| 3 | New York | United States | 18,008 | Basel | Switzerland | 28,971 |
| 4 | Eindhoven | Netherlands | 16,373 | Bridgeport | United States | 24,294 |
| 5 | Albany | United States | 10,805 | Washington | United States | 11,016 |
| 6 | Arnhem | Netherlands | 9,625 | Tokyo | Japan | 10,945 |
| 7 | Trenton | United States | 9,346 | Dallas | United States | 9,124 |
| 8 | Los Angeles | United States | 8,175 | Munich | Germany | 5,810 |
| 9 | Umea | Sweden | 5,360 | Cologne | Germany | 5,508 |
| 10 | Berlin | Germany | 4,501 | Dublin | Ireland | 4,408 |
| 11 | Pittsburgh | United States | 4,127 | Geneva | Switzerland | 4,391 |
| 12 | Cambridge | United Kingdom | 4,025 | Seattle | United States | 3,082 |
| 13 | Hamburg | Germany | 3,904 | Helsinki | Finland | 2,355 |
| 14 | Philadelphia | United States | 3,665 | Osaka | Japan | 2,055 |
| 15 | San Antonio | United States | 3,300 | Mumbai | India | 2,027 |

| 16 | Jacksonville | United States | 3,132 | Zurich | Switzerland | 2,014 |
|----|---|---|---|---|---|---|
| 17 | Austin, Texas | United States | 2,872 | Chicago | United States | 1,553 |
| 18 | Brussels | Belgium | 2,789 | Houston | United States | 1,519 |
| 19 | Vasteras | Sweden | 2,507 | Taipei City | Taiwan | 1,198 |
| 20 | Norwich | United States | 2,453 | Oxnard | United States | 1,039 |
| 21 | Toronto | Canada | 2,260 | Minneapolis | United States | 1,037 |
| 22 | Beijing | China | 2,252 | Heerlen | Netherlands | 1,033 |
| 23 | Stuttgart | Germany | 2,081 | Delhi | India | 1,012 |
| 24 | Frankfurt | Germany | 2,014 | Stockholm | Sweden | 923 |
| 25 | Bangalore | India | 1,950 | London | United Kingdom | 813 |
| 26 | Santa Cruz | United States | 1,926 | Suwa | Japan | 714 |
| 27 | Vienna | Austria | 1,905 | Providence | United States | 529 |
| 28 | Auckland | New Zealand | 1,875 | Midland, Michigan | United States | 469 |
| 29 | Uppsala | Sweden | 1,697 | Johannesburg | South Africa | 460 |
| 30 | Montreal | Canada | 1,534 | Luxembourg | Luxembourg | 388 |
| 31 | Pune | India | 1,449 | Gothenburg | Sweden | 329 |
| 32 | Greenville | United States | 1,294 | Malmö | Sweden | 240 |
| 33 | Taoyuan City | Taiwan | 1,268 | New Haven | United States | 235 |
| 34 | Shanghai | China | 1,237 | Findlay | United States | 215 |
| 35 | Cleveland | United States | 1,236 | Milwaukee | United States | 214 |
| 36 | Cincinnati | United States | 1,140 | Memphis | United States | 206 |
| 37 | Nuremberg | Germany | 998 | Atlanta | United States | 204 |
| 38 | Jamshedpur | India | 996 | Graz | Austria | 175 |
| 39 | Haifa | Israel | 992 | Niles-Benton Harbor | United States | 171 |
| 40 | Villach | Austria | 977 | Davenport | United States | 159 |
| 41 | Rochester | United States | 976 | Corning | United States | 156 |
| 42 | Canton | United States | 962 | Copenhagen | Denmark | 152 |
| 43 | Korla City | China | 925 | Salt Lake City | United States | 145 |
| 44 | Siena | Italy | 923 | Aarhus | Denmark | 144 |
| 45 | Detroit | United States | 921 | Seoul | South Korea | 137 |
| 46 | Marietta | United States | 917 | Mirny | Russia | 83 |
| 47 | Syracuse | United States | 832 | Kingsport | United States | 76 |
| 48 | Manchester-Nashua | United States | 810 | St. Louis | United States | 76 |
| 49 | Lyon | France | 790 | Yingtan | China | 69 |
| 50 | Shunan | Japan | 773 | Hartford | United States | 66 |

It needs mentioning that Scandinavian cities occupy a much better position as cities hosting corporate R&D, in contrast to Dublin and Amsterdam (see, Table 6). In the past two decades, Dublin has become the home of many high-tech and pharmaceutical companies originally headquartered in the United States (for example, Eaton, Ingersoll-Rand, Seagate Technology), but which have recently registered their head office in Dublin, taking advantage of its low corporate tax rate. However, in most cases, these companies' R&D facilities, even the companies themselves remain in the United States. Amsterdam's position is almost exclusively influenced by the presence of Royal Philips: the company is headquartered in Amsterdam, but its largest research centre, which is responsible for 60 percent of its articles, is located in Eindhoven.

## 5 Conclusion

In her pioneer work, Saskia Sassen (1991; 2001) defines global cities as the most important command and control centres in the world economy and as sites of production, including the production of innovation, in leading industries. That is, on the one hand, global cities are home to headquarters of the largest MNCs in the world, and on the other hand, they host the majority of the R&D activities of MNCs. In this paper, we put Sassen's theory to the test by conducting a bibliometric analysis. According to our hypothesis, not only are the largest MNCs headquartered in the global cities, but as their role as the most innovative areas in the world, the cities also host the R&D-oriented subsidiaries, branches, and corporate research centres of domestic and foreign MNCs. In order to confirm this hypothesis, we compared the number of scientific articles about the *Forbes* 2000 companies in the Scopus database with respect to two approaches: in the first case, we assigned all articles written about MNCs to headquarters' cities (home city approach), while in the second case, we assigned articles to cities where the research was conducted, and the articles were created (host city approach). In light of our hypothesis, New York, London, Tokyo, and Paris ranked at the top of the hierarchy by having the largest number of scientific articles by either approach.

The results show that the cities of New York and Tokyo excel far above the others. New York, in particular, seems to be the most important node for global corporate R&D activities while Tokyo primarily hosts the corporate R&D activities of globally influential domestic companies. London's position among the global cities is undoubtable based on both approaches, but Paris has a better position as a city of headquarters (3rd position) than a city hosting corporate R&D activities (5th position). We have found two cities that may have a leading role as nodes of global corporate R&D activities in the future. Both of them are homes to the world's fastest growing industries: San Jose is the international centre of the information technology industry, and Boston is the leading global hub of the biotechnology industry. In both rankings, only one city appears from the emerging economies, namely Beijing; which is still one of the leading command and control centres in the world economy. Moreover, its role as a major node of corporate R&D activities is predictable (see, for example, Zhou, 2005; Liefner et al. 2006; Andersson et al. 2014).

**Acknowledgement**

This paper is supported by the **János Bolyai Research Scholarship** of the Hungarian Academy of Sciences.